\documentclass[
    reprint,
    superscriptaddress,
    amsmath,
    amssymb,
    aps,
    prx,
    nofootinbib
]{revtex4-2}

\usepackage{textcomp}  % Additional text symbols
\usepackage{graphicx}  % Include figure files
\usepackage{hyperref}  % Add hypertext capabilities
\usepackage{siunitx}
\usepackage{xspace}
\usepackage{tabularx}
\usepackage[symbol]{footmisc}
\newcommand{\ax}{$a_x=579\,$nm}
\newcommand{\ay}{$a_y=1187\,$nm}

\newcommand{\PRL}{Phys. Rev. Lett.}
\newcommand{\NaturePhysics}{Nat. Phys}
\newcommand{\PRX}{Phys. Rev. X}
\newcommand{\PRA}{Phys. Rev. A}
\newcommand{\NaturePhotonics}{Nat. Photonics}

\makeatletter
\def\maketitle{
\@author@finish
\title@column\titleblock@produce
\suppressfloats[t]}
\makeatother

\begin{document}

\newcommand{\partitle}[1]{\section{#1}}

\newcommand{\papertitle}{Continuous operation of large-scale atom arrays in optical lattices}

\title{\papertitle{}}

 \author{Flavien Gyger}%\thanks{These authors contribute equally to this work.}
     \affiliation{Max-Planck-Institut f\"{u}r Quantenoptik, 85748 Garching, Germany}
     \affiliation{Munich Center for Quantum Science and Technology (MCQST), 80799 Munich, Germany}
     
 \author{Maximilian Ammenwerth}%\thanks{These authors contribute equally to this work.}
     \affiliation{Max-Planck-Institut f\"{u}r Quantenoptik, 85748 Garching, Germany}
     \affiliation{Munich Center for Quantum Science and Technology (MCQST), 80799 Munich, Germany}

 \author{Renhao Tao}%\thanks{These authors contribute equally to this work.}
     \affiliation{Max-Planck-Institut f\"{u}r Quantenoptik, 85748 Garching, Germany}
     \affiliation{Munich Center for Quantum Science and Technology (MCQST), 80799 Munich, Germany}
   \affiliation{Fakultät für Physik, Ludwig-Maximilians-Universit\"{a}t, 80799 Munich, Germany}

 \author{Hendrik Timme}
     \affiliation{Max-Planck-Institut f\"{u}r Quantenoptik, 85748 Garching, Germany}
     \affiliation{Munich Center for Quantum Science and Technology (MCQST), 80799 Munich, Germany}

 \author{Stepan Snigirev}
     \affiliation{PlanQC GmbH, 85748 Garching, Germany}

 \author{\\Immanuel Bloch}
     \affiliation{Max-Planck-Institut f\"{u}r Quantenoptik, 85748 Garching, Germany}
     \affiliation{Munich Center for Quantum Science and Technology (MCQST), 80799 Munich, Germany}
     \affiliation{Fakultät für Physik, Ludwig-Maximilians-Universit\"{a}t, 80799 Munich, Germany}

 \author{Johannes Zeiher}
 \email{johannes.zeiher@mpq.mpg.de}
     \affiliation{Max-Planck-Institut f\"{u}r Quantenoptik, 85748 Garching, Germany}
     \affiliation{Munich Center for Quantum Science and Technology (MCQST), 80799 Munich, Germany}
     \affiliation{Fakultät für Physik, Ludwig-Maximilians-Universit\"{a}t, 80799 Munich, Germany}

\date{\today}

\begin{abstract}
    Scaling the size of assembled neutral-atom arrays trapped in optical lattices or optical tweezers is an enabling step for a number of applications ranging from quantum simulations to quantum metrology.
    However, preparation times increase with system size and constitute a severe bottleneck in the bottom-up assembly of large ordered arrays from stochastically loaded optical traps.
    Here, we demonstrate a novel method to circumvent this bottleneck by recycling atoms from one experimental run to the next, while continuously reloading and adding atoms to the array.
    Using this approach, we achieve densely-packed arrays with more than $1000$ atoms stored in an optical lattice, continuously refilled with a net $2.5$ seconds cycle time and about $130$ atoms reloaded during each cycle.
    Furthermore, we show that we can continuously maintain such large arrays by simply reloading atoms that are lost from one cycle to the next.
    Our approach paves the way towards quantum science with large ordered atomic arrays containing thousands of atoms in continuous operation.\end{abstract}
\maketitle

%%%%%%%%%%%%%%%%%%%%%%%%%%%%%%%%%%%%%%%%%%%%
%               Introduction               %
%%%%%%%%%%%%%%%%%%%%%%%%%%%%%%%%%%%%%%%%%%%%
\section{Introduction}
Atom arrays stored in optical lattices or optical tweezers are a promising platform for quantum simulation, quantum computation, and quantum metrology~\cite{Saffman2016, Browaeys2020, kaufman2021, Daley2022, Morgado2021, Katori2011}.
A usual experimental sequence to control atoms in optical lattices or optical tweezers starts with the preparation of an ensemble, followed by the simulation, calculation or metrology sequence.
Finally, a destructive measurement of the state of the system is performed that typically renders a recycling of atoms from one cycle to the next impossible.
The subsequent preparation of a fresh ensemble of atoms requires significantly more time than the actual experimental sequence, leading to a dead time that becomes significantly longer for large arrays.
This naturally calls for a different mode of operation in which only the lost atoms are prepared and replaced in each cycle.
While recently demonstrated in a seminal work in bulk gases~\cite{Chen2022}, a re-use of atoms and cyclic operation with microscopic control are challenging and require non-destructive detection in combination with resorting to replenish lost atoms~\cite{Endres2016,Barredo2016}.
High-fidelity and low-loss detection of single atoms is now routinely achieved for several species in optical lattice and optical tweezers~\cite{Gross2017,Covey2019,Saskin2019,tao2023,young2023,Blodgett2023}.
Although first steps towards extended operation of atom arrays have recently been demonstrated in small-scale systems with finite reservoirs~\cite{Pause2023, norcia2023}, truly continuous operation requires schemes for reloading new atoms that do not affect the atomic array already present in the system.
Effective strategies to ``hide" stored atoms during the reloading of new atoms have recently been demonstrated in dual-element arrays of two alkali-atom species~\cite{singh2022,Singh2023}, where atoms of one species are only minimally affected by forming a magneto-optical trap (MOT) of the other element, thereby enabling continuous operation with arrays of each element prepared in alteration.
An alternative route is offered by utilizing the more complex level structure available in alkaline earth(-like) atoms such as strontium or ytterbium.
Here, two separate optical series with different total spin and metastable states exist, which has proven useful for a variety of applications in combination with microscopic control~\cite{Young2020,Jenkins2022,Ma2022,Urech2022,Schine2022,Lis2023,Ma2023,Scholl2023,norcia2023,Eckner2023,shaw2023,Shaw2024}.
In particular, the metastable states can also be used to effectively hide stored atoms while forming a MOT for the ground-state atoms.
Although this level structure has been shown to be well adapted for the preparation of one-dimensional atomic arrays with near-unity filling based on dark-state enhanced loading combined with site control using an acousto-optic deflector (AOD)~\cite{shaw2023}, continuous loading has so far remained an elusive goal.
\begin{figure*}[t!]
    \centering
    \includegraphics[width=\textwidth]{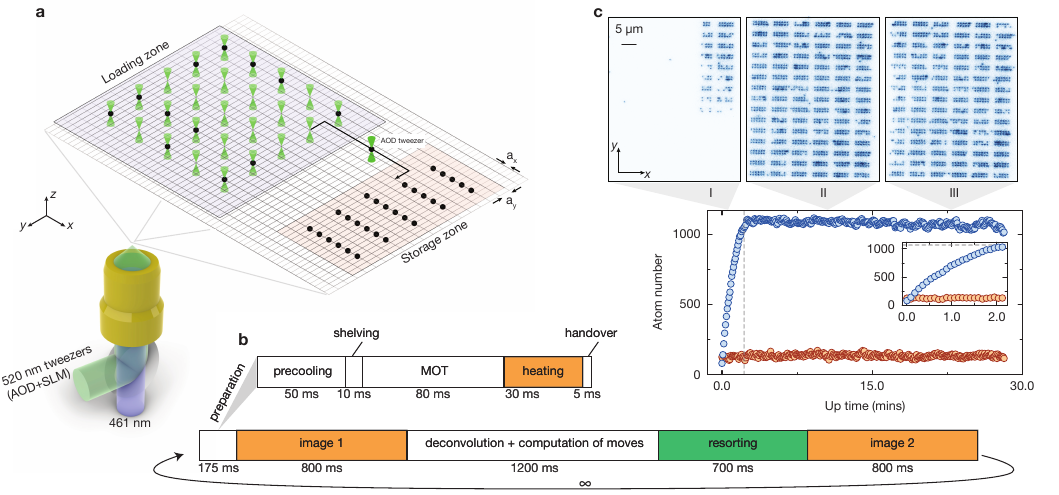}
    \caption{
    \textbf{Concept and demonstration of continuous operation.} \textbf{a} Main features of our experimental machine. We use a $1040$\,nm stationary bow-tie optical lattice as our physics array (gray grid with \ax\ and \ay). We subdivide the accessible area into a loading zone and a storage zone. The loading zone is overlapped with a stationary tweezer array at $520\,$nm. Atoms are transported with AODs that steer a single beam in the lattice plane from the loading zone to the storage zone.
    \textbf{b} Experimental sequence of our continuous loading scheme.
    \textbf{c} Exemplary single shots at various instances in time of an iteratively assembled array exceeding $1000$ atoms on average. Bottom graph: Atom number of the continuously operated array (blue) and atom number in the loading area (red). Inset: Zoom-in of the atom number in the build-up phase.
    }
    \label{fig:1}
\end{figure*}

%%%%%%%%%%%%%%%%%%%%%%%%%%%%%%%%%%%%%%%%%%%%
%               Here, we show...           %
%%%%%%%%%%%%%%%%%%%%%%%%%%%%%%%%%%%%%%%%%%%%
Here, we show a novel scheme that combines several of the foregoing aspects to realize continuously operated large-scale atom arrays with atom numbers continuously exceeding $1000$ atoms and reaching up to $1247$ atoms.
Our scheme relies on a continuously operated storage zone in an optical lattice, which is periodically replenished from a loading zone and a MOT.
Using a bichromatic combination of loading and storage arrays, we achieve excellent spatial control over the loading zone, strongly suppressing loading of sites in the storage register.
Loading about $130$ new atoms for each cycle, we grow and then continuously maintain an array of atoms in an optical lattice with more than $1000$ atoms, which is about eight times larger than the number of atoms loaded during each cycle.
Our results mark a paradigm shift in the operation of quantum simulators and quantum computers based on neutral atoms to iteratively assembled and continuously operated arrays.

%%%%%%%%%%%%%%%%%%%%%%%%%%%%%%%%%%%%%%%%%%%%
%         Describe the main idea           %
%%%%%%%%%%%%%%%%%%%%%%%%%%%%%%%%%%%%%%%%%%%%
\section{Assembly of large arrays}
The architecture of our apparatus~\cite{tao2023} for continuous operation is shown in Fig.~\ref{fig:1}.
We operate the experiment in one region spanning about $130~\mu\text{m}\times130~\mu$m, corresponding to the area that the AODs can currently address.
This region is centered above our objective lens and contains about $24000$ trapping sites in a bow-tie lattice.
We divide this lattice region into two sub-regions: a loading zone and a storage zone; see Fig.~\ref{fig:1}a.
The loading zone of the lattice is replenished from a reservoir of $323$ tweezers, which are overlapped in three dimensions with the lattice sites. These tweezers are themselves loaded with $^{88}$Sr atoms from a dual-stage MOT, on the broad $^1$S$_0$-$^1$P$_1$ transition at $461\,$nm and on the narrow-line $^1$S$_0$-$^3$P$_1$ transition at $689\,$nm~\cite{xu2003}.
For high-fidelity detection, we transfer the atoms from the tweezer array into the optical lattice~\cite{young2022, Schine2022, young2023, tao2023} and perform fluorescence imaging therein.
Our cyclic sequence is presented in Fig.~\ref{fig:1}b and has a cycle time of $2.5\,$ seconds, excluding data processing.

In the first iteration, we load in average $N_L$ atoms from the MOT into the tweezer array.
Then, we transfer the atoms in the lattice and perform a high-fidelity and low-loss imaging to detect the position of loaded atoms in the loading zone~\cite{tao2023}.
Detected atoms are subsequently displaced on demand from the loading zone to the storage zone by a moving tweezer controlled by crossed AODs.
After resorting, we shelve the atoms in the storage zone to the long-lived metastable $^3$P$_0$-state and subsequently refill the tweezer array in the loading zone from a MOT created at the location of our lattice.
Shelving in the magnetically-insensitive clock state protects the stored atoms from loss occurring during the MOT.
A subsequent fluorescence image of both the storage and the loading zone depumps the atoms in $^3$P$_0$ back to the ground-state $^1$S$_0$, revealing unoccupied sites in the storage zone that need to be refilled.
The possibility of reusing atoms across experimental runs results in an important scaling advantage for the achievable array sizes.
The largest array size is reached when the number of atoms lost during the previous cycle is precisely balanced by the number of atoms replenished in the current cycle.
% From one cycle to the next, only the atoms lost in the last cycle have to be replenished with the available atoms from a reservoir.
%
This condition limits the maximally reachable atom number to $N_\infty = \beta N_L$, where the amplification factor
\begin{equation}
    \label{eq:1}
    \beta = \frac{1-\alpha_r}{\alpha_c}
\end{equation}
is proportional to the atom move success probability $1-\alpha_r$ and inversely proportional to the cycle loss $\alpha_c$, which quantifies the fraction of stored atoms that are lost from one cycle to the next.
The effective number of atoms that can be added to the array in each cycle is $N_{L,\mathrm{eff}} = N_L\cdot (1-\alpha_r)$ which is smaller than the loaded atom number due to atom loss during the transport.
For typically achieved cycle losses in this work ($\alpha_c\approx 10\%$), the saturated atom numbers already significantly exceed the number of loaded atoms per cycle $N_L$; see Fig.~\ref{fig:1}d.
We note that, in principle, the loading zone could be fully overlapped with the storage zone by interleaving loading sites and storage sites.
Such a configuration would be beneficial, since it increases the available space for the storage area and shortens the move distance between the loading and storage zone.
However, in our configuration, the tweezer light at $520\,$nm induces considerable loss of atoms stored in the $^3$P$_0$ states and prevents us from overlapping the two zones. We find evidence that photoionization of the $^3$P$_0$ state is responsible for this loss, see Appendix~\ref{sec:tw_ionisation}~\cite{Ma2023}, which could be mitigated in the future at alternative tweezer wavelengths, such as $813\,$nm.
In the following, we perform a detailed characterization of all steps involved in the continuous operation of our arrays.
%%%%%%%%%%%%%%%%%%%%%%%%%%%%%%%%%%%%%%%%%%%%
%               Shelving + heating         %
%%%%%%%%%%%%%%%%%%%%%%%%%%%%%%%%%%%%%%%%%%%%

\begin{figure}[t!]
    \centering
    \includegraphics{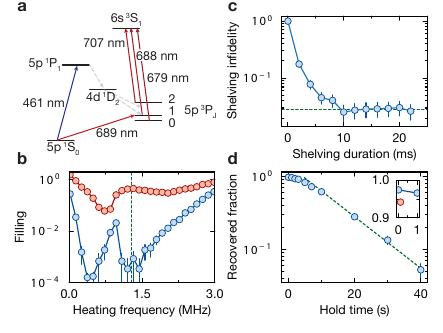}
    \caption{
    \textbf{Shelving and trap-selective heating.} \textbf{a} Energy level of $^{88}$Sr.
    \textbf{b} Resulting filling after a trap-selective $689\,$nm Sisyphus-heating pulse applied to atoms trapped in the lattice (blue) and in the combined tweezer-lattice potential (red) as a function of the applied frequency detuning. The trace is normalized to the loading fraction without heating pulse for the lattice and normalized to our tweezer number ($323$) for the combined potential. An extinction of $5\times 10^{-4}$ for ground-state lattice atoms is achieved when choosing a frequency detuning of $1.28$\,MHz (green dashed line) while the atoms in the combined tweezer-lattice potential remain almost unaffected.
    \textbf{c} Round-trip shelving infidelity as a function of shelving duration, reaching $3\%$ (green dashed line) after $10\,$ms.
    \textbf{d} Shelving lifetime in our lattice at $200\,\mu$K, reaching $13$ seconds, extracted from an exponential fit to the last four data points (green dashed line). Inset: Zoom-in of the hold time region below one second. The red dot indicates our actual recovered fraction when a MOT is created within the hold time.
    }
    \label{fig:2}
\end{figure}
\section{Reloading the reservoir}%
An important step in the cyclic operation of the array concerns the transfer of atoms from $^1\text{S}_0$ to $^3\text{P}_0$ before reloading the reservoir.
Population shelving is currently implemented via a combination of light at $689\,$nm and $688\,$nm driving $^1$S$_0\rightarrow^3$P$_1\rightarrow^3$S$_1$.
An additional repumper beam at $707\,$nm renders the $^3$P$_0$-state the only dark state; see Fig.~\ref{fig:2}a.
The fraction of shelved atoms reaches $97\,\%$ after $10\,$ms of pumping, as shown in Fig.~\ref{fig:2}c. 
We note that a higher shelving fraction could be achieved by a combination of coherent shelving and incoherent pumping, which would directly reduce our cycle loss~\cite{madjarov2020}.
At our lattice trap depth of $200\,\mu$K, the lifetime of atoms in $^3$P$_0$-state is $13\,$s; see Fig.~\ref{fig:2}d. % limited by off-resonant scattering of the lattice light~\cite{dorscher2018}. <--- given the shape of the trace, I am not fully sure about this statement. It might be temperature-limited.
The lifetime significantly exceeds the total duration during which the shelved atoms stay in $^3$P$_0$ of $115\,$ms in our sequence, such that holding of shelved atoms by itself induces an insignificant loss.
However, we observe a small increase of the shelved atom loss induced by the presence of the MOT, resulting in a total of $6\%$ shelving loss during the $^3$P$_0$ preparation and atom reloading steps; see Fig.~\ref{fig:2}d inset.
Importantly, we make our MOT without a repumper at $679\,$nm, which would otherwise deplete the shelved atoms in $^3$P$_0$.
This considerably reduces the number of loaded atoms.
Still, with optimized parameters, we obtain a filling fraction as high as $40\%$ atom loading probability in the loading tweezer-lattice register sites after parity projection.
An additional consequence of the absence of a $^3$P$_0$ repumper is the decay of a small fraction of atoms into $^3$P$_0$ via $^1$D$_2\rightarrow^3$P$_{2}\rightarrow^3$S$_1$, where they become indistinguishable from the shelved atoms in the stored array.
To counteract this effect, the $^1$D$_2$-state could be repumped to higher-lying $^1$P$_1$-states on transitions either at $716\,$nm or $448\,$nm~\cite{Samland2023}, which would increase the MOT loading fraction and remove defects originating from accidentally shelved atoms in $^3$P$_0$ during the MOT stage.

Importantly, the MOT loads the entire lattice, including the loading zone and the storage zone.
To remove ground-state atoms from the lattice everywhere except from the sites overlapped with the tweezer array in the loading zone, we subsequently apply an essential trap-selective heating pulse.
This heating pulse is optimized to remove ground-state atoms in the lattice, while leaving both ground-state atoms in the tweezer array and $^3$P$_0$ atoms in the lattice intact; see Fig.~\ref{fig:2}b. 
To realize this selective removal of atoms, we use a beam at $689\,$nm tuned to a repulsive Sisyphus heating regime.
This heating feature is unique to transitions narrow enough to spatially resolve the differential trap depth between the $^1$S$_0$-state and the $^3$P$_1$-state.
For the chosen detuning, a net kinetic energy gain is realized between subsequent excitation-decay cycles, and thus leads to a fast, highly parallel, and well-controllable heating mechanism of ground-state atoms in the lattice.
Importantly, the atoms in the loading zone, which experience a combined bichromatic lattice-tweezer potential, are shielded from the heating resonance and thus remain trapped; see Fig.~\ref{fig:2}b.
Consequently, our selective heating pulse effectively removes all atoms in $^1$S$_0$ that only experience the lattice potential with an extinction of more than $5\times10^{-4}$.
This prevents uncontrolled storage of atoms directly loaded from the MOT in the storage zone.
%
%This heating pulse doesn't actually leave the atoms in the tweezers unaffected. Instead, it performs light-assisted collisions leading to parity projection~\cite{Schlosser2001}.
%
%We note that even in a monochromatic trap, a similar trap-selective heating pulse could be implemented by exploiting its harmonic confinement.
%
Alternatively, site-selective parallel addressing can also be used to hide already loaded sites from being further loaded, as effectively shown in 1d~\cite{shaw2023}, with direct extensions using state-selective parallel addressing in higher dimensions~\cite{Zhang2023}.
%
%After the heating pulse, the atoms in the tweezers are transferred into the lattice by a tweezer potential ramps down followed by a turn off of the tweezer light.
%
%
%Now we take a lattice image to reveal the location of the newly loaded atoms in our loading region, and also the vacancies in our target storage atoms.
%

%
\section{Resorting in an optical lattice}
\begin{figure}
    \centering
    \includegraphics[scale=1]{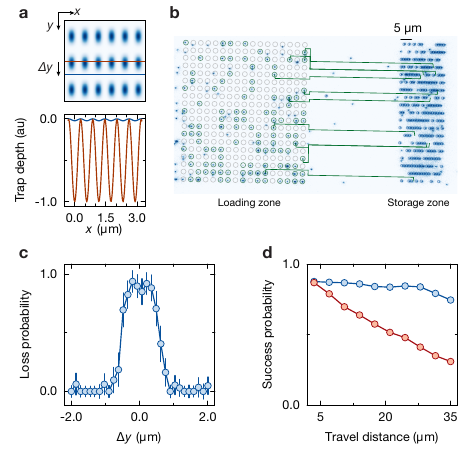}
    \caption{
    \textbf{Atom rearrangement.} \textbf{a} Upper graph: Energy landscape of our folded lattice. Lower graph: Reduction of the trap depth modulation experienced by an atom traveling in between the lattice sites (blue line in the upper graph), as compared to through the lattice sites (red line in the upper graph).
    \textbf{b} Resorting algorithm favoring horizontal moves along corridors in between the lattice sites.
    \textbf{c} Atom loss probability as a function of the distance between an occupied lattice site and the traveling resorting tweezer. Disturbance leading to atom loss are observed below 1 $\mu$m of distance.
    \textbf{d} Atom move success probability $1-\alpha_r$ as a function of the traveling distance, when the moves are performed through the lattice sites (red points) and when they are performed in between the lattice sites (blue points). The atom move success probability is defined as the probability of not losing the atom during the entire move operation.
    }
    \label{fig:3}
\end{figure}
\begin{figure*}
    \centering
    \includegraphics{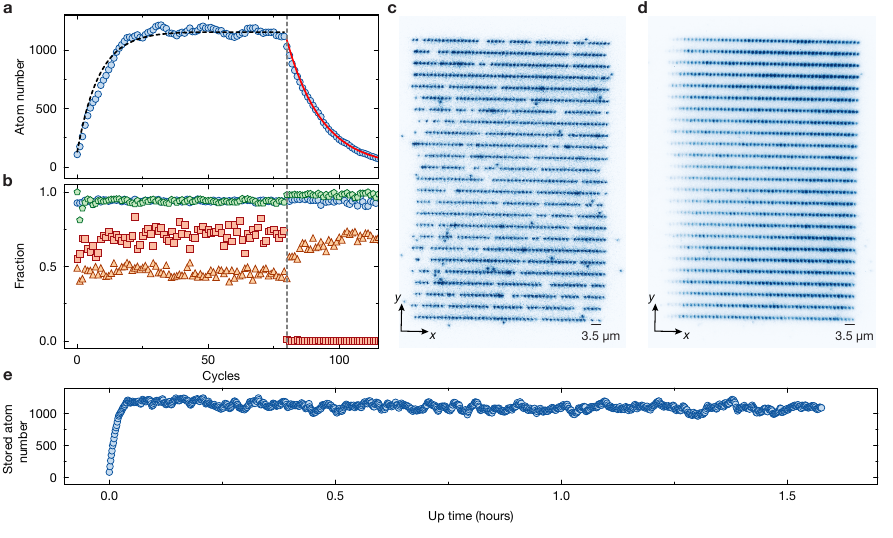}
    \caption{
    \textbf{Characterization of continuous operation.}
    \textbf{a} Number of atoms in the storage area as a function of cycles.
    After 80 cycles, we disable the resorting and let the array naturally decay.
    The black dashed line is our model from Eqs.~\eqref{eq:2} with measured average parameters, while the red line is an exponential fit that enables us to measure the cycle loss $\alpha_c$ (neglecting the influence of resorting in this case).
    \textbf{b} Evolution of the loading fraction (orange triangles), resorting move success probability (red squares), shelved survival fraction (blue circles) and stored survival fraction during resorting (green pentagons) as a function of cycles.
    \textbf{c} Single-shot image of the storage array containing 1230 atoms.
    \textbf{d} Average image of the storage array.
    \textbf{e} Continuous operation of the array for more than 1.5 hour.
    }
    \label{fig:4}
\end{figure*}
The next important step in our cyclic sequence is the rearrangement of newly loaded atoms to vacancies in the target of the sorted array.
We perform such moves using a pair of AODs similar to previous work~\cite{de2019,Eckner2023, young2023, Bluvstein2024}.
In contrast to the case of optical tweezer arrays, atoms moving through lattice sites experience a large periodic trap depth modulation, potentially leading to strong heating. 
In the special case of a bow-tie lattice, this modulation can be strongly reduced by moving atoms in between the lattice sites, thereby ameliorating any heating effects; see Fig.~\ref{fig:3}a.
In line with this expectation, we directly observe that long moves in between the lattice sites exhibit a significantly higher success probability than moves through lattice sites; see Fig.~\ref{fig:3}d.
%
%, i.e. the probability of finding an atom after the move on the desired target site in the optical lattice,
In this measurement and during the continuous loading operation, all moves are performed with a peak velocity of $54\,\mu$m$/$ms.
In particular, for the long moves considered in this work, movement between sites is crucial.
Our lattice has spacings between sites of \ax\, and \ay\, along the $x$- and $y$-axis respectively, and is therefore particularly suited for horizontal moves; see Fig.~\ref{fig:3}a.
To optimally leverage the favorable geometry of our lattice, our resorting procedure has been designed to predominantly move between lattice sites with a five-stroke move pattern; see Fig.~\ref{fig:3}b.
The first stroke removes the atom out of its lattice site and brings it in between lattice sites.
The second stroke displaces the atom outside of the loading area.
The third and fourth strokes adjust the vertical ($y$-axis) and horizontal ($x$-axis) positions respectively to almost match that of its final destination, and finally the last stroke inserts the atom in its final location; see Fig.~\ref{fig:3}b.
%
%In this work, the resorting operation only moves atoms from the loading zone to the storage zone.
%
%So far, no attempt is made to correct any of the defects accumulating in the storage array.
%
Each move consists of (i) a slow ramp-up of the moving tweezer potential depth to about ten times the lattice depth to extract the target atom out of the lattice, (ii) a sequence of parameterized frequency chirps encoding velocity profiles and turns, and (iii) a final ramp-down of the tweezer depth to release the atom in the desired target lattice site in the storage zone.
The initial and final intensity ramp durations are $400\,\mu$s each, to ensure adiabaticity~\cite{weitenberg2011}.

The imperfect resorting process affects continuous loading in two ways:
First, it reduces the effective atom number $N_{L,\mathrm{eff}}$ that can be maximally added to the storage array in each cycle when atoms are lost during transfer.
Second, by traveling at a close distance to an already stored atom, the unintended perturbation of the trapping site can result in a loss of the stored atom and thus directly increases the cycle loss.
We observe a corresponding limit for the minimum distance between a stored atom and the trajectory of the moving tweezer, which is approximately equal to $1\,\mu$m; see Fig.~\ref{fig:3}c.
This minimum distance sets a limit for the minimal achievable spacing between atoms in the storage register.
The total resorting duration for each cycle is about $700\,$ms. This duration could be drastically reduced by implementing a more complex parallel resorting scheme~\cite{Endres2016,ebadi2021,tian2023,young2023}.
%
%Moreover, at each cycle, the tweezer array is actively displaced in the horizontal plane to follow the free-running lattice drift with a feed-forward.
%
%Since the spacings, angle and horizontal shifts of both traps are identical, they spatially overlap.
\section{Continuous operation}
Finally, combining all steps, we demonstrate the ability to build and maintain a large-scale, densely packed tweezer array for more than an hour; see Fig.~\ref{fig:4}.
After the initial loading stage, the number of atoms stored in the array remained above $1000$ atoms for most of the operation time.
From the atom loss within and across different cycles, we can extract both the cycle loss $\alpha_c$, as well as the resorting loss $\alpha_r$ that enter in a simple model for the build-up and saturation of the stored atom number $N_i$ at each cycle $i$,
\begin{align}
\label{eq:2}
%\begin{aligned}
    N_{i+1} &= (1-\alpha_c)\cdot N_i + (1-\alpha_r)\cdot N_L,\qquad i = 0, 1, \ldots \nonumber \\
    N_\infty &= \frac{(1-\alpha_r)\cdot N_L}{\alpha_c}.
%\end{aligned}
\end{align}
The dashed line in Fig.~\ref{fig:4}a is computed from these equations using time-average values of the measured cycle loss and resorting loss, and is in excellent agreement with the measured atom number.
Four parameters of interest are extracted from the occupation matrices and plotted in Fig.~\ref{fig:4}b: 
(i) The loading fraction, defined as the number of loaded atoms in the loading zone $N_L$ normalized to total number of tweezer sites ($323$); (ii)
the resorting move success probability $1 -\alpha_r$, defined as the number of loaded atoms that are successfully moved into the storage zone, normalized to the total loaded atom number $N_L$; (iii) the shelved survival fraction, defined as the number of atoms that survive the shelving, holding and repumping operation between two cycles, normalized to the total atom number in the storage zone; and (iv) the stored atom survival fraction during resorting, defined as the number of atoms in the stored array that survive the resorting procedure without being moved themselves, normalized to the total atom number in the storage zone.
We observe that, in our system, both the shelved survival fraction and the stored atom survival fraction during resorting contribute equally to the cycle loss.
After 80 cycles, we disable the resorting operation and let the stored array decay. 
Subsequently, we observe that (i) the loading curve rises, since the loaded atoms are no longer removed and accumulate in the loading zone; and (ii) the stored atoms survival fraction during resorting rises, since no resorting operation disturbs the stored array.

A more detailed correlation analysis relating the measured final atom numbers to the extracted cycle loss, the resorting infidelity and the reloaded atom number, directly reveals that the fluctuations of the final atom number are most strongly correlated with the cycle loss, see Appendix~\ref{sec:appendix_b}. Such a behavior is expected from the inverse scaling of steady-state atom number with cycle loss, and highlights the large potential gains in a further optimized sequence.
% 
%%%%%%%%%%%%%%%%%%%%%%%%%%%%%%%%%%%%%%%%%%%%
%  		   Conclusion & Outlook            %
%%%%%%%%%%%%%%%%%%%%%%%%%%%%%%%%%%%%%%%%%%%%
\section{Conclusion and Outlook}
To conclude, we have presented a first realization of densely packed, continuously loaded atom arrays stored in an optical lattice. 
Prospectively, we expect that our technique could allow for the assembly of significantly larger atom arrays than what we have demonstrated.
In previous work, the efficiency of shelving atoms in $^3$P$_0$ has been shown to reach $99.7\%$~\cite{madjarov2020}.
For a clock-state lifetime of the order of $100\,$s with a reduced lattice potential~\cite{dorscher2018} and a typical MOT stage of $100\,$ms duration, the shelved atoms loss during the MOT stage could be reduced to $0.1\%$, leading to a total shelving loss as low as $0.4\%$.
Assuming our measured vacuum lifetime of $273\,$s, the typical vacuum-limited loss for a 1-second cycle time experiment is roughly also equal to $0.4\%$, leading to a total cycle loss of about $\alpha_c \approx 0.8\%$
Together with an achieved atom loss due to resorting moves as low as $\alpha_r \approx 2\%$~\cite{young2023}, Eq.~\eqref{eq:1} predicts the achievable amplification factor to reach $\beta > 100$.
With such a large amplification, one could reach about $10000$ atoms in a single array with $100$ loaded atoms at each cycle, provided that a sufficient area for storage and high-fidelity detection is available~\cite{tao2023}.
Deterministically loaded arrays~\cite{Brown2019,Jenkins2022}, or directly loaded lattices~\cite{tao2023} as loading zones could further boost the achievable steady-state atom numbers potentially by orders of magnitude.
We want to emphasize that, compared to directly assembling such a large number of atoms in a single experimental cycle, continuous loading has the benefit of only moving the newly loaded atoms for each cycle, thus reducing both move-induced losses as well as the resorting time overhead by the amplification factor $\beta$.
Moreover, maintaining coherence during the reload stage, for example for atoms placed in a specially shielded physics array, would open up exciting new perspectives for quantum metrology or quantum information tasks~\cite{Singh2023,norcia2023}.
Such large, continuously maintained atom arrays combined with the recently demonstrated fast, high-fidelity quantum gates~\cite{evered2023} and elementary logical quantum circuits~\cite{Bluvstein2024} make neutral atoms a promising platform for quantum computing and quantum simulation at scale.

\emph{Note added:} During the preparation of the manuscript, we became aware of related work, where similar results
have been reported using arrays of $^{171}$Yb~\cite{norcia2024}.

\begin{acknowledgments}
% \section*{Acknowledgments}
We acknowledge funding by the Max Planck Society (MPG) the Deutsche Forschungsgemeinschaft (DFG, German Research Foundation) under Germany's Excellence Strategy--EXC-2111--390814868, from the Munich Quantum Valley initiative as part of the High-Tech Agenda Plus of the Bavarian State Government, and from the BMBF through the programs MUNIQC-Atoms and MAQCS.
This publication has also received funding under Horizon Europe programme HORIZON-CL4-2022-QUANTUM-02-SGA via the project 101113690 (PASQuanS2.1).
J.Z. acknowledges support from the BMBF through the program “Quantum technologies - from basic research to market” (SNAQC, Grant No. 13N16265).
M.A. and R.T. acknowledge funding from the International Max Planck Research School (IMPRS) for Quantum Science and Technology. M.A acknowledges support through a fellowship from the Hector Fellow Academy.
F.G. acknowledges funding from the Swiss National Fonds (Fund Nr. P500PT\textunderscore203162).

\end{acknowledgments}

%%%%%%%%%%%%%%%%%%%%%%%%%%%%%%%%%%%%%%%%%%%%
%  		   APPENDIX            %
%%%%%%%%%%%%%%%%%%%%%%%%%%%%%%%%%%%%%%%%%%%%
\appendix
\section{$^3$P$_0$ ionization in $520\,$nm tweezers}
\label{sec:tw_ionisation}
\begin{figure}[h]
    \centering
    \includegraphics{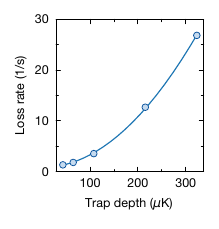}
    \caption{
    \textbf{$^3$P$_0$ ionization from $520$ nm tweezers.}
    Measured and parabolic fit of $^3$P$_0$ lifetime as a function of the tweezers trap depth. The extracted quadratic contribution to the loss rate is $250$ s$^{-1}$ (mK)$^{-2}$.
    }
    \label{fig:S1}
\end{figure}

In our experiments, we find the $^3$P$_0$ lifetime of atoms in $520$\,nm tweezers to be limited to $40\,$ms for tweezers at a trap depth of $300\,\mu$K, and to depend quadratically on the trap depth, as shown in Fig.~\ref{fig:S1}.
This suggests that light at $520$\,nm could ionize atoms stored in $^3$P$_0$ in a two-photon excitation.
In our continuous loading experiment, we observed that the shelving lifetime was considerably decreased for atoms in the loading zone, even when atoms were trapped in a lattice site not overlapping with a tweezer.
As a consequence of this observation, we spatially separated our loading and storage zone to mitigate this extra tweezer-induced cycle loss.
We anticipate that using another wavelength for the tweezers (e.g.\ $813$\,nm) would sidestep this issue and increase the available space for both the loading and storage zones.

\section{Correlation analysis on the atom number fluctuation in the stored array}
\label{sec:appendix_b}

\begin{figure}[t!]
    \centering
    \includegraphics{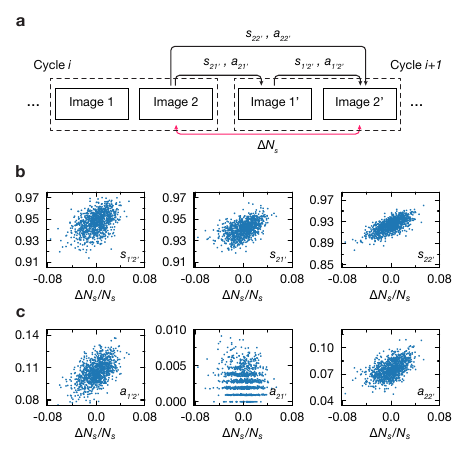}
    \caption{
    \textbf{Correlations analysis on atom number fluctuation.}
    \textbf{a} Sequence of images taken during continuous loading for the current cycle $i$ and the next cycle $i+1$ indicating between which images various quantities of interest are computed at the cycle $i$; see text for more details.
    \textbf{b} Survival fraction of atoms from one image to another, as a function of atom number fluctuation.
    \textbf{c} Gain fraction of atoms from one image to another, as a function of atom number fluctuation.
    }
    \label{fig:S2}
\end{figure}

To understand the origin of the atom number fluctuation in continuous operation of the stored array, we study the correlations between the fluctuations and a set of parameters based on the atom occupation during various stages of the sequence.
In this analysis, we only consider atoms placed at sites of the stored array defined by a custom target; all atoms which are not in a site of the stored array, even if they are placed in the storage zone (defects), are ignored.
We define the survival fraction $s_{mn}$ as the sum over all stored array sites that are both filled in image $m$ and image $n$, normalized to the occupation of image $m$.
Therefore, $0 \leq s_{mn}\leq 1$ quantifies the atoms in the stored array that survived from image $m$ to a succeeding image $n$.
Similarly, we define the gain fraction $a_{mn}$ as the sum over all stored array sites that are empty in image $m$ but filled in image $n$, normalized to the occupation of image $n$.
Therefore, $0 \leq a_{mn} \leq 1$ quantifies the atoms that have appeared in the stored array from image $m$ to a succeeding image $n$.
For the survival fraction $s_{mn}$ and gain fraction $a_{mn}$, we label the images using an augmented image index $m,n = \{1, 2, 1', 2'\}$, where the indices $1$ and $2$ refer respectively to the first and second image of a cycle $i$, and where the indices $1'$ and $2'$ refer respectively to the first and second image of the next cycle $i+1$; see Fig.~\ref{fig:S2}a.
The fluctuations are quantified by the quantity $\Delta N_s/N_s$, where $N_s$ is the number of filled sites in our stored array while $\Delta N_s$ is defined as the difference of the number of filled sites in the stored array between two consecutive cycles.
The correlations of the atom number fluctuations $\Delta N_s/N_s$ with $s_{mn}$ and $a_{mn}$ are summarized in Fig.~\ref{fig:S2}b,c. 
We quantify the strength of the correlations using the Pearson coefficient $\rho(X, Y)$ between two random variables $X$ and $Y$, defined as
\begin{equation}
    \rho(X, Y) = \frac{\mathrm{cov}{(X, Y)}}{\sigma{(X)}\sigma{(Y)}},
\end{equation}
where $\mathrm{cov}$ is the covariance and $\sigma{(X)}$, $\sigma{(Y)}$ are the standard deviations of $X$ and $Y$ respectively.
We observe that $\Delta N_s/N_s$ correlates most strongly with the cycle survival ($s_{22'}$), comprising the contribution of the shelved survival fraction ($s_{21'}$) and the stored atom survival fraction ($s_{1'2'}$); see Table~\ref{table:pearson}.

\begin{table}[t!]
\vspace{12pt}
\centering
\begin{tabular}{ |c|c c c c c c| } 
 \hline
 Quantity & $s_{1'2'}$ & $s_{21'}$ & $s_{22'}$ & $a_{1'2'}$ & $a_{21'}$ & $a_{22'}$ \\
 \hline
 Pearson coefficient & 0.46 &  0.50 & 0.68 & 0.53 & -0.12 & 0.51 \\
 \hline
\end{tabular}
\caption{Pearson coefficients of the correlated quantities.}
\label{table:pearson}
\end{table}

Interestingly, we also observe a non-vanishing value of $a_{21'}$, despite no atoms being intentionally added to the array between the second image of the cycle $i$ and the first image of the cycle $i+1$.
Since neither imperfect heating nor atoms accidentally shelved in $^3$P$_0$ during the MOT yield large enough contributions to explain the observed increased atom number from one cycle to the next, we speculate this effect to mostly arise from misclassification of our deconvolution algorithm.

\bibliography{ContinuousLoading}

\end{document}